\documentclass[pre,twocolumn,groupedaddress,floatfix]{revtex4-1}
\usepackage{amssymb,amsmath}
\usepackage{graphicx}
\usepackage{subfigure}
\usepackage[english]{babel} 
\usepackage{float}
\usepackage{color}

\usepackage{lipsum}
\begin{document}
\newcommand{\be}{\begin{equation}}
\newcommand{\ee}{\end{equation}}
\newcommand{\rojo}[1]{\textcolor{red}{#1}}

\title{The fractional saturable impurity}

\author{Mario I. Molina}
\affiliation{Departamento de F\'{\i}sica, Facultad de Ciencias, Universidad de Chile, Casilla 653, Santiago, Chile}

\date{\today }

\begin{abstract} 
We examine analytically and numerically the effect of fractionality on a saturable bulk and surface impurity embedded in a 1D lattice. We use a fractional Laplacian introduced previously by us, and by the use of lattice Green functions we are able to obtain the bound state energies and amplitude profiles, as a function of the fractional exponent $s$ and saturable impurity strength $\chi$ for both, surface and bulk impurity. The transmission is obtained in closed form as a function of $s$ and $\chi$, showing
strong deviations from the standard case, at small fractional exponent values. The selftrapping of an initially-localized excitation is qualitatively similar for the bulk and surface mode, but in all cases 
complete confinement is obtained at $s\rightarrow 0$, as shown theoretically and observed numerically.

\end{abstract}

\maketitle

{\bf 1.\ Introduction}.\\ 
When a defect is inserted in a discrete, periodic system such as a chain of atoms or an optical waveguide array, the original translational symmetry is broken and causes that one of the states detaches from the band and form a localized mode centered at the impurity position. It has been proven that, for 1D and 2D lattices there is always a localized bound state centered at the impurity\cite{slater,harrison}, regardless of the strength of the impurity. The rest of the modes remain extended but they are no longer sinusoidal. The single-defect system is the starting point for the study of the more complex system with a finite fraction of disorder, where the main phenomenon of study is Anderson localization\cite{economou1,economou2}. Some examples of linear impurities include junction defects between two optical or network arrays\cite{miro}, coupling defects, discrete networks for routing and switching of discrete optical solitons\cite{christo}, and also in simple models for magnetic metamaterials, modeled as periodic arrays of split-ring resonators, where magnetic energy can be trapped at impurity positions\cite{wang}.

When nonlinearity is added to a periodic system, mode localization and self-trapping of energy can occur. This localized mode which exists in this nonlinear but otherwise completely periodic system is known as a discrete soliton. In most cases this concentration of energy on a small region increases with the nonlinearity strength and, as a consequence, the nonlinear mode becomes effectively decoupled from the rest of the lattice. In the high nonlinearity limit, the effective nonlinearity is concentrated in a small region around the soliton, and thus, we can consider the rest of the lattice as approximately linear. We are then left with a linear lattice containing a single nonlinear impurity. This simplified system is easier to treat theoretically and closed-form solutions are sometimes possible. In  condensed matter, nonlinear impurities appear when one dopes a material with atoms or molecules that have strong local couplings. In optics, the system of interest is a dielectric waveguide array, where one of the guides is judiciously doped with an element with strong polarizability. A more recent example is magnetic metamaterials, where the  system is an array of inductively coupled split-ring resonators, where a linear/nonlinear impurity ring is obtained by, for instance,  inserting a linear/nonlinear dielectric inside its slit to change its resonance frequency. In the absence of the impurity, the modes are  extended magneto-inductive plane waves, and when a capacitive impurity is introduced, a localized mode is created.  The nonlinear impurity concept has also been explored in studies of embedded solitons\cite{malomed}.

A common approach when dealing with generic impurities is to make an educated guess about the shape of the mode (usually exponential) which then leads to the mode energy and exact spatial profile. However, this procedure might work only partially. For one thing, in the presence of nonlinearity, the number of modes depends on the available energy content, and there are possible bifurcation separating different modes with different stabilities.  Also, when boundaries are involved, like impurities close to a surface, the need for a more formal treatment is apparent. An elegant method for dealing with impurity problems is the technique of lattice Green functions\cite{green, barton,duffy}. Originally devised for linear problems, it has been shown that it can also be extended to nonlinear cases\cite{molina2,molina3,molina4,molina5}.  This is the method we will follow in this work, with the added feature of fractionality.

The concept of fractionality has gained considerable interest in recent years. Roughly speaking, it consists on a generalization of the standard derivative of integer order by one of fractional order. 
It all started with the correspondence between Leibnitz and L'Hopital about possible generalizations of the concept of an integer derivate. The starting point was the calculation of $d^{s} x^{k}/ dx^{s}$, for  $s$ a real number. This means
\be 
{d^{n} x^k\over{d x^n}}= {\Gamma(k+1)\over{\Gamma(k-n+1)}} x^{k-n} \rightarrow {d^s  x^k\over{d x^s }} = {\Gamma(k+1)\over{\Gamma(k-s +1)}} x^{k-s }.\label{eqx}
\ee
where $\Gamma(x)$ is the Gamma function. From Eq.(\ref{eqx}) the fractional derivative of an analytic function $f(x)=\sum_{k} a_{k} x^{k}$ can be computed by deriving the series  term by term. However, this simple procedure is not exempt from ambiguities. These early studies were followed later by rigorous work by several mathematicians including Euler, Laplace, Riemann, and Caputo to name some, and  converted  fractional calculus from a mathematical curiosity into a research field of its own\cite{fractional1,fractional2,fractional3,hilfer}. Several possible definitions for the fractional derivative are known, each one with its own advantages and disadvantages. One of the most common definitions is the Riemann-Liouville form
\be
\left({d^{s }\over{d x^{s }}}\right) f(x) = {1\over{\Gamma(1-s )}} {d\over{d x}} \int_{0}^{x} {f(s)\over{(x-s)^{s }}} ds
\ee
another common form, is the Caputo formula
\be
\left({d^{s }\over{d x^{s }}}\right) f(x) = {1\over{\Gamma(1-s )}} \int_{0}^{x} {f'(s)\over{(x-s)^{s }}} ds
\ee
where, $0<s <1$ is the fractional exponent. This formalism that extends the usual integer calculus to a fractional one, with its definitions of a fractional integral and fractional 
derivative, has found application in several fields: fractional kinetics and anomalous diffusion\cite{metzler,sokolov,zaslavsky}, fluid mechanics\cite{fluid2}, strange kinetics\cite{shlesinger}, Levy processes in quantum mechanics\cite{levy}, fractional quantum mechanics\cite{laskin,laskin2}, plasmas\cite{plasmas}, electrical propagation in cardiac tissue\cite{cardiac}, epidemics\cite{epidemics} and biological invasions\cite{invasion}.\\

In this work we will study the effect of fractionality on the bound state and plane-wave transmission properties of a saturable impurity, seeking to characterize these properties as a function of the fractional exponent $s $.

\noindent
{\bf 2.\ The model}.\\

Let us consider a generic excitation propagating along a 1D chain periodic chain that contains a single saturable impurity at site $d$:
\be
i\ {d C_{n}\over{d t}} + V (C_{n+1} + C_{n-1}) + \delta_{n,d}\,\chi {C_{n}\over{1+|C_{n}|^2}}=0\label{eq1}
\ee
where $C_{n}(t)$ is the probability amplitude for finding the excitation at site $n$ at time $t$, V is the hopping parameter, and $\chi$ is the nonlinear parameter. 
In an optical context\cite{milutin}, Eq.(\ref{eq1}) describes an array of semiconductor (GaAs/AlGaAs) optical waveguides where one of the guides is doped with a photorefractive element such as lithium niobate doped with a metal: $\mbox{Fe: LiNbO}_{3}$. 
The term $V(C_{n+1} + C_{n-1})$ is basically the discrete Laplacian $\Delta_{n} C_{n}=C_{n+1}-2 C_{n}+C_{n-1}$. Then, eq.(\ref{eq1}) can be cast as 
\be
i {d C_{n}\over{d t}} + 2 V C_{n} + V \Delta_{n} C_{n} + \delta_{n,d}\,\chi {C_{n}\over{1+|C_{n}|^2}}=0.\label{eq2}
\ee
The next step is to replace the discrete Laplacian $\Delta_{n}$ by its fractional form 
 $(\Delta_{n})^s $ in Eq.(\ref{eq2}). The form of this fractional discrete laplacian has been found in closed form, and is given by\cite{discrete laplacian}:
\be
(-\Delta_{n})^s  C_{n}=\sum_{m\neq n} K^s (n-m) (C_{n}-C_{m}),\hspace{0.5cm}0<s <1 \label{delta}
\ee
where
\be
K^s(m)={4^s \Gamma(s +(1/2))\over{\sqrt{\pi}|\Gamma(-s )|}}{\Gamma(|m|-s )\over{\Gamma(|m|+1+s )}}.\label{eq7}
\ee
where $\Gamma(n)$ is the Gamma function and $s $ is the fractional exponent. When $s =1$ the system reduces to the standard one with an integer Laplacian. We look for stationary modes $C_{n}(t) = \phi_{n}\ \exp(i \lambda t)$, obtaining a system of nonlinear difference equations for $\phi_{n}$:
\begin{eqnarray}
(-\lambda+2V)\ \phi_{n}& +& V\sum_{m\neq n}K^s (n-m)(\phi_{m}-\phi_{n})+\nonumber\\
                       &  & + \ \delta_{n,d}\,\chi\ {\phi_{n}
\over{1+|\phi_{n}|^2}}=0\label{eq9}
\end{eqnarray}
where, without loss of generality, the $\phi_{n}$ can be chosen as real. Also the $2 V$ term in Eq.(\ref{eq9}) must be replaced by $V$ when $n$ corresponds to any of the two edge sites, for a finite chain.

In the absence of the saturable impurity, we have solutions of the type $C_{n}=\phi_{n}\, \exp(i k n)$. After inserting this ansatz into Eq.(\ref{eq9}), we obtain the dispersion relation 
\begin{widetext}
\be
\lambda(k) = 2V - 4 V \sum_{m=1}^{\infty} K^s (m) \sin((1/2) m k)^2\label{dispersion}
\ee
or, in closed form
\be
\lambda(k)=2 V - {16 V\ \Gamma(s +(1/2))\over{\sqrt{\pi}\ \Gamma(1+s )}}\Big( 1-\exp(-i k)\ s \ \Gamma(1+s )[\ R(1,1-s ,2+s ;\exp(-i k))+\exp(2 i k) \ R(1,1-s ,2+s ;\exp(i k))\ ] \Big)\label{dispersion2}
\ee
\end{widetext}
where $R(a,b,c;z)={}_2 F _{1}(a,b,c;z)/\Gamma(c)$ is the regularized hypergeometric function. 

{\em Bulk impurity}.\ In the presence of the impurity ($\chi\neq 0$), it becomes easier to compute its properties by using the formalism of lattice Green function rather than working directly from Eq.(\ref{eq9}). In our case the Hamiltonian can be cast as
\be
H = H_{0} + H_{1}
\ee
\begin{eqnarray}
H_{0} &=& \sum_{n} \epsilon_{n}|n\rangle\langle n|\, + \sum_{n,m} |m\rangle V_{n,m}\,\langle n|\\
 H_{1} &=&  {\chi\over{1+|\phi_{d}|^2}} |d\rangle\langle d|\label{H1}
\end{eqnarray}
with
\be
\epsilon_{n} = 2 V - V \sum_{m\neq n} K^s (n-m)
\ee
and
\be
V_{n m}=K^s (n-m)=V_{m n}
\ee
and the Dirac notation has been used. Hamiltonian $H_{0}$ is the unperturbed Hamiltonian, that is, the Hamiltonian in the absence of the impurity, while $H_{1}$ is the perturbation due to the presence of the saturable impurity at $n=d$, where for the bulk impurity $d$ is far away from the boundaries of the lattice, while for the surface impurity $d=0$. 
The equations of motion for the amplitudes $C_{n}$ are given by $i\,d C_{n}/dt=\partial H/\partial C_{n}^{*}$. The Green function is defined as
\be
G(z) = {1\over{z-H}}.\label{G}
\ee 
It can be written in a more explicit form as
\be
G_{n m}^{(0)}(z)={1\over{2 \pi}} \int_{-\pi}^{\pi} {e^{i k (n-m)} dk\over{z-\lambda(k)}}\label{G0}
\ee
where $n$ and $m$ are lattice positions, and the notation $G_{m n}^{(0)} = \langle m|G^{(0)}|n \rangle$ has been used and where $\lambda(k)$ is given by Eq.(\ref{dispersion2}). Now we formally expand Eq.(\ref{G}) as a perturbation series in $H_{1}$:
\be
G(z) = G^{(0)} + G^{(0)}\ H_{1}\  G^{(0)} + G^{(0)}\ H_{1}\ G^{(0)}\ H_{1}\  G^{(0)} \cdots
\ee
After resuming the perturbative series to all orders, we obtain
\be
G(z) = G^{(0)} + {\epsilon \over{1 - \epsilon\, G_{d d}^{(0)}}}  G^{(0)}|d\rangle \  \langle d| G^{(0)},
\ee
where, 
\be
\epsilon\equiv {\chi\over{(1+|\phi_{d}|^2)}}.\label{epsilon}
\ee
According to the general theory\cite{green}, the energy $z_{b}$ of the bound state is given by the poles of $G_{d d}(z)$ 
\be
1 = \epsilon\ G_{d d}^{(0)}(z_{b})=\chi {G_{d d}^{(0)}(z_{b})\over{1+|\phi_{d}^{(b)}|^2}},
\ee
while the square of the mode amplitude at site $n$ is given by the residue of $G_{n m}(z)$ at the pole
\be
|\phi_{n}^{(b)}|^2 = -{{G_{n d}^{(0)}}^2 (z_{b})\over{G_{d d}'^{(0)}(z_{b})}}.
\label{phin2}
\ee
Also, from general Eqs. (\ref{phin2}) and (\ref{G0}), it can be easily proven that the bound state mode is always normalized: $\sum_{n} |\phi_{n}^{(b)}|^2 = 1$. The bound state energy equation becomes
\be
{1\over{\chi}} = {G_{d d}^{(0)}(z_{b}) G_{d d}^{'(0)}(z_{b})\over{G_{d d}^{'(0)}(z_{b})-G_{d d}^{(0)2}(z_{b})}}.\label{energy}
\ee
This formalism is also useful to compute the transmission of plane waves across the impurity, The transmission amplitude is given by $T\sim 1/|1-\epsilon G_{d d}^{+}(z)|^2$, while the reflection amplitude is given by $R\sim \epsilon^2 |G_{d,d}^{+}(z)|^2/|1-\epsilon G_{d d}^{+}(z)|^2$, with $G^{+}(z) = \lim_{\eta\rightarrow 0} \,G(z+i \eta)$ and $\epsilon$ given by Eq.(\ref{epsilon}). After normalizing theses amplitudes by $N=T+R$ the transmission coefficient $t$ and reflection coefficient $r$ can be expressed as
\begin{eqnarray}
t(z) &=& {1\over{1 + \epsilon^2 |G_{d d}^{+}(z)|^2}}\label{t}\\
r(z) &=&{\epsilon^2 |G_{d d}^{+}(z)|^2\over{1 + \epsilon^2 |G_{d d}^{+}(z)|^2}}, \label{r}
\end{eqnarray}
and $z$ is inside the band $\lambda(k)$. Since the transmission coefficient is also equal to the probability at the impurity site, $\epsilon=\chi/(1+t)$, Eq.(\ref{t}) becomes a cubic equation for $t$:
\be
(1+t)^2-t (b+(1+t)^2)=0 \ \ \ \mbox{with}\ \ \  b\equiv \chi^2\ |G_{d d}^{+}(z)|^2 \label{eq:b}
\ee
with real solution
\begin{widetext}
\be
t = {1\over{3}}\left(  -1 - {2^{1/3} (-4+3 b)\over{(\ 16+9 b + \sqrt{27 b\,(32+b(-13+4 b))}\ \ )^{1/3}}}+
2^{-1/3} (\ 16+9 b + \sqrt{27 b\,(32+b(-13+4 b))}\ \ )^{1/3} \right).\label{trans}
\ee
\end{widetext}
Fractionality is implicit in Eq.(\ref{trans}) through $b$ which depends on $G^{+}$ which, in turn, depends on $\lambda(k)$ through Eq.(\ref{dispersion2}).

{\em Surface impurity}.\ In the case of the surface impurity, located at $n=0$, we have to take into account the presence of the boundary. That is, since there is no lattice to the left of $n=0$, $G_{m n}^{(0)}$ should vanish identically at $n=-1$. Thus, $G_{m n}^{(0)}=G_{m n}^{\infty}-G_{m, -n-2}^{\infty}$
where $G_{m n}^{\infty}$ is the unperturbed Green function for the infinite lattice. Using the representation (\ref{G0}), we have
\be
G_{m n}^{(0)}= {1\over{2 \pi}} \int_{-\pi}^{\pi} {e^{i k (m -n)}\over{z-\lambda(k)}}-
{1\over{2 \pi}} \int_{-\pi}^{\pi} {e^{i k (m+n+2)}\over{z-\lambda(k)}}. \label{images}
\ee
The computation of the bound state energy and bound state amplitude proceed as before, using this new $G_{m n}^{(0)}$ (\ref{images}), extracted from the method of images. 

{\bf Results}.\\  

We begin by taking a look at the bound state energy equation (\ref{energy}). Figure 1 shows the RHS of Eq.(\ref{energy}) as a function of the frequency $z$. The horizontal dashed line represents the value of some $1/\chi$, whose intersection with the RHS curve give us the bound state energy. For the bulk case, the RHS curve diverges at the edges of the band and there is a single bound state solution for any $\chi$. Also, as the fractional exponent is decreased, the lower band edge shrinks causing the negative energy eigenvalue to shift towards less negative values. For the surface impurity case, the situation is similar to the bulk case, except that now the RHS of Eq. (\ref{energy}) does not diverge at the band edges, which means that a minimum $\chi$ value must be reached in order for an intersection with $1/\chi$ to occur. The behavior with a change in fractional exponent $s$ is similar to the bulk case.
\begin{figure}[t]
 \includegraphics[scale=0.225]{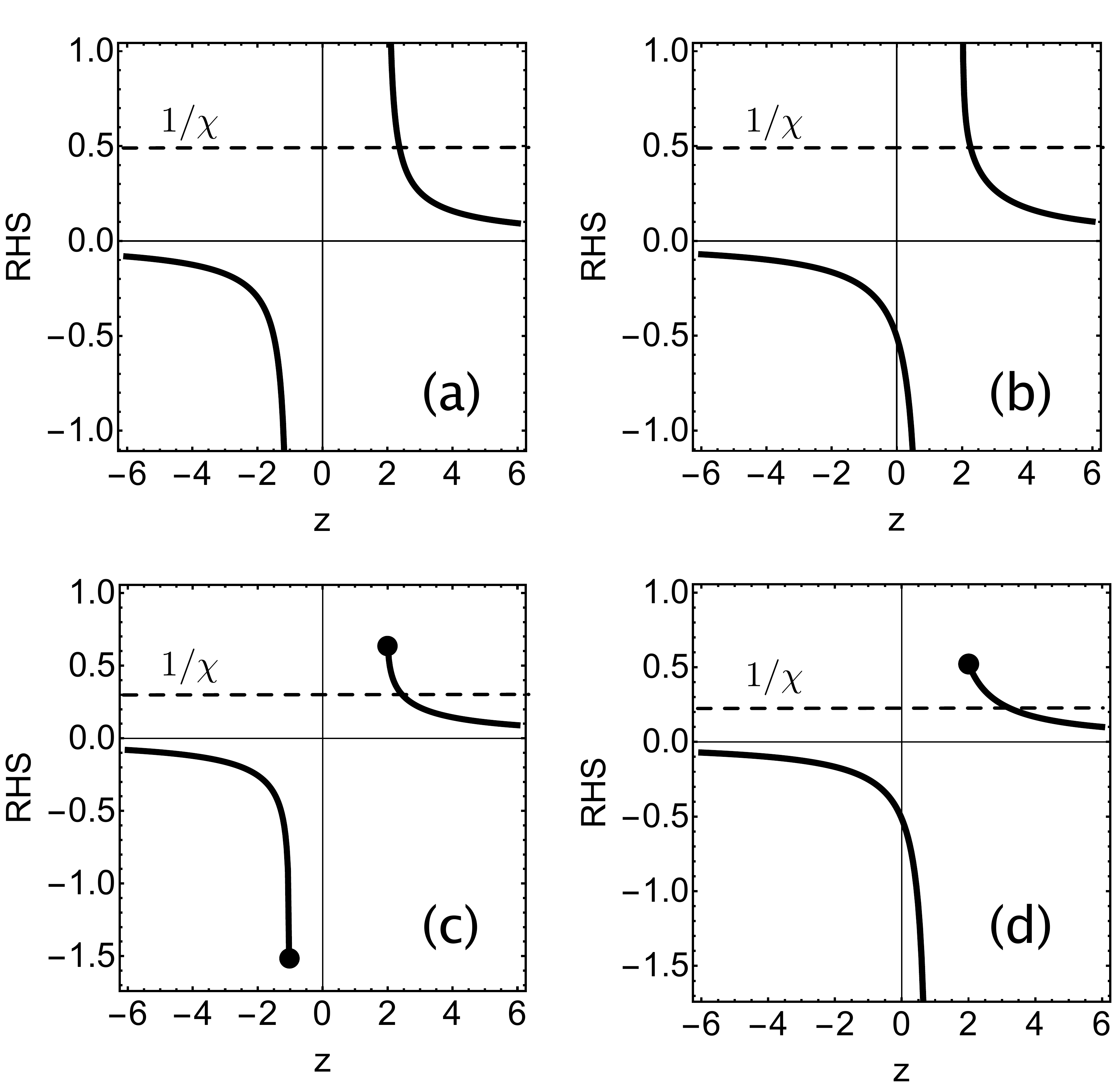}
  \caption{Bound state condition (Eq.(\ref{energy})) for the bulk impurity (a) $s=0.8$ and (b) $s=0.2$, and the surface impurity (c)$s=0.8$ and (d)$s=0.2$. For the surface impurity, the dots denotes the value of the RHS at the band edges. The intersection of the horizontal line with the RHS gives the bound state energy.
}  \label{fig1}
\end{figure}
Figure 2 shows the bulk and surface bound state energies as a function of the impurity strength for several fractional exponents. In all cases there is a single bound state mode that lives outside the bands, whose width decreases with a decrease en $s$. The bandwidth is given by $\lambda(0)-\lambda(\pi)$, where $\lambda(k)$ is given by Eq.(\ref{dispersion2}). 
In the limit of large impurity strength, it is possible to obtain the asymptotic value for $z_{b}$: Since $z_{b}$ is monotonic with $\chi$, we have that $G_{0 0}^{(0)}(z_{b})\rightarrow 1/z_{b}, G'^{(0)}_{0 0}(z_{b})\rightarrow 1/z_{b}^2$. After replacing this into the bound state energy equation (\ref{energy}), we obtain $z_{b}\rightarrow \chi/2$ and $|\phi_{n}|^2 \rightarrow \delta_{n,d}$, i.e., complete localization at the impurity site.  
\begin{figure}[t]
 \includegraphics[scale=0.21]{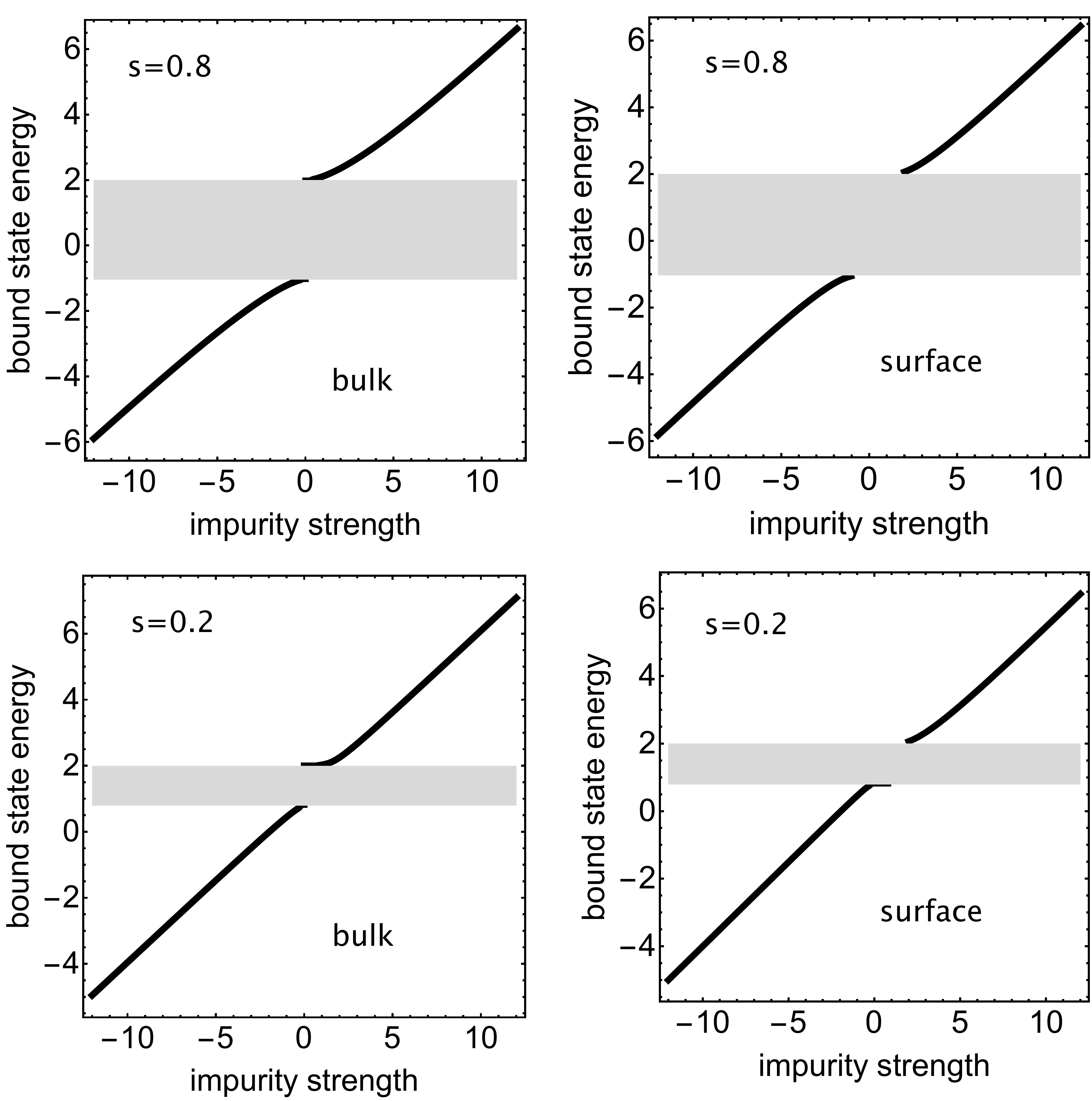}
  \caption{Bound state energy versus impurity strength for the bulk and surface impurity cases, and for a couple of different fractional impurities $s$.
}  \label{fig2}
\end{figure}
\begin{figure}[h]
 \includegraphics[scale=0.15]{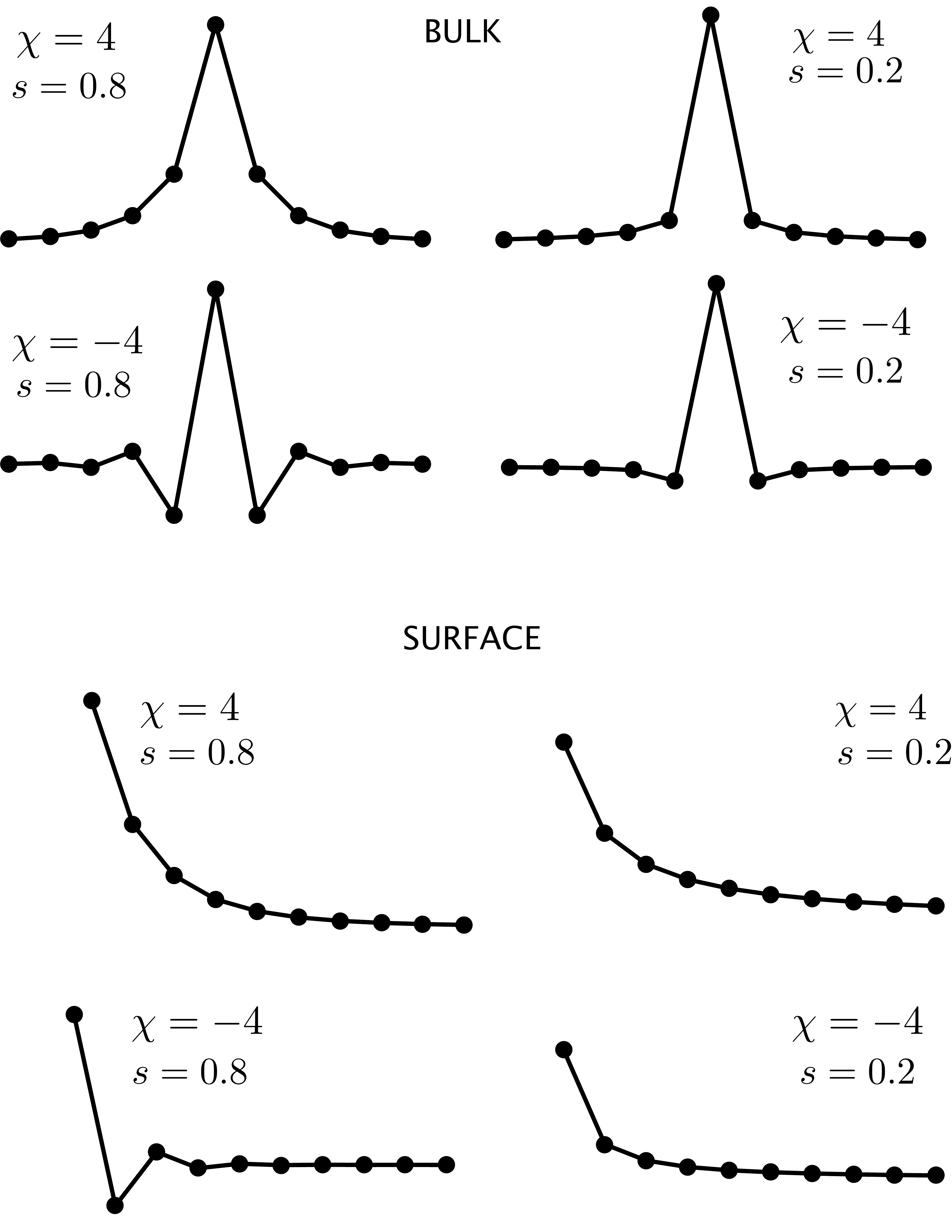}
  \caption{Bulk and surface bound state profiles for some impurity strengths and fractional exponents.
}  \label{fig3}
\end{figure}
\begin{figure}[h]
 \includegraphics[scale=0.40]{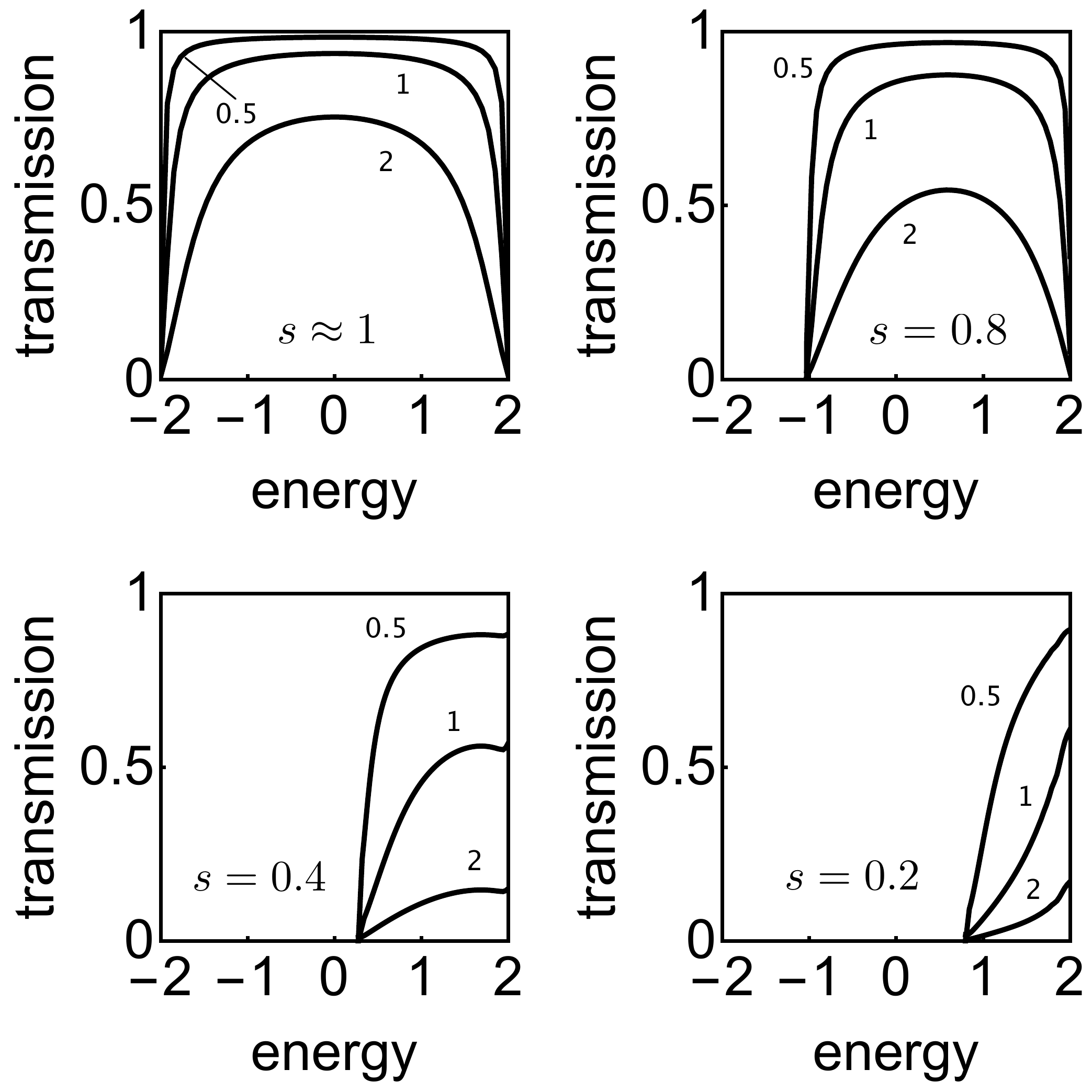}
  \caption{Transmission of plane waves across the saturable impurity, for different fractional exponents $s$ and different impurity strengths $\chi$ labelling each curve.
}  \label{fig4}
\end{figure}
We notice that the phenomenology is, in general, similar to the case of a linear impurity\cite{linear}. This could be explained by the nature of the saturable nonlinearity where one could express the nonlinearity as $\chi_{\mbox{eff}}=\chi/(1+|\phi_{d}|^2)<\chi$. Thus, the saturable impurity is always ``weaker'' than the linear one.

Figure 3 shows some bulk and surface bound state profiles for several values of the impurity strengths and fractional exponents. They look similar to the localized modes of cubic impurities. However, 
while for the standard case $(s\approx 1)$ the change $\chi\rightarrow -\chi$ produces an staggered version of the mode, $\phi_{n}\rightarrow (-1)^n \phi_{n}^{}$, here for $s < 1$ this  symmetry is debilitated, specially at low $s$ values where the staggered mode is lost completely. Same happens for the surface bound state.

Figure 4 shows the transmission of plane waves across the saturable impurity, as a function of energy. We use the exact expression Eq.(\ref{trans}).  For each fractional exponent $s$, we plot curves corresponding to different impurity strengths $\chi$. The first thing we notice is that, as $s$ decreases, the energy range for transmission shrinks. This is due to the reduction of the bandwidth with $s$ and was already observed in Fig.\ref{fig2}. On the other hand, for each fixed $s$, the transmission decreases with increasing $\chi$, as expected. It can also be proved that there are no resonances ($t=1$): If we set $t=1$ in Eq.(\ref{eq:b}), we obtain the condition $b=0$. However, $b = \chi^2\ |G_{0 0}^{+}|^2 > 0$, so there is no resonance except in the limit of no impurity. Finally, let us consider the issue of selftrapping. We put an initial excitation on the impurity site and observe its dynamical evolution at long times. When there is a finite 
\begin{figure}[t]
 \includegraphics[scale=0.17]{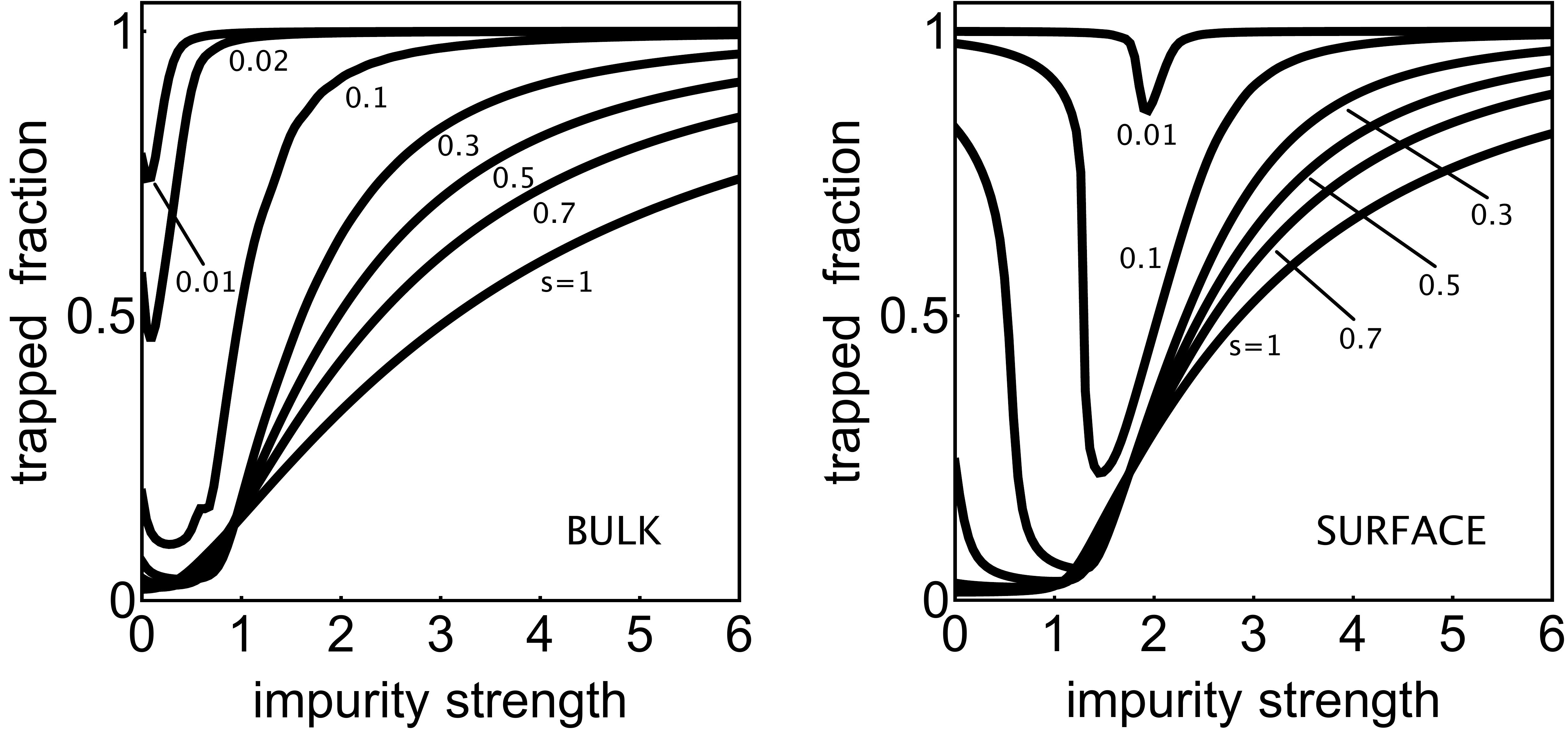}
  \caption{Time-averaged trapped fraction at impurity site versus nonlinearity strength for several fractional exponents $s$, for the bulk and surface impurity. The numbers labelling each curve represent the fractional exponents.
}  \label{fig4}
\end{figure}
portion remaining on the impurity site, we speak of selftrapping. This type of selftrapping is a common feature of many discrete nonlinear lattices for a variety of nonlinearity types.
Figure 5 shows the selftrapping for the bulk and surface impurity, and for several values of the fractional impurity $s$. Roughly speaking both cases are qualitatively similar, with an amount of trapped fraction that increases with a decrease in $s$. A very distinctive feature of Fig.5 is the presence, in both cases, of selftrapping at small value of $\chi$. This `linear' trapping increases with a decrease in $s$ and at $s\rightarrow 0$, the trapped fraction approaches unity, independent of the impurity strength. This behavior can be explained as follows: When $s\rightarrow 0$, $K^{s}(m) \rightarrow (s/m) + {\cal O}(s^2)$. Thus, evolution equation (\ref{eq2}) reduces to
\be
i {d C_{n}\over{d t}} + 2 V C_{n} + \delta_{n,d}\,\chi {C_{n}\over{1+|C_{n}|^2}}=0.\label{eqa}
\ee
with initial condition $C_{n}(0)=\delta_{n d}$, where $d=0$ for the surface impurity, or $d\sim N/2$ for the bulk impurity. After multiplying Eq.(\ref{eqa}) by $C_{n}^{*}$, we have
\be
i C_{n}^{*} {d C_{n}\over{d t}} + 2 V |C_{n}|^2 + \delta_{n,d}\,\chi {|C_{n}|^2\over{1+|C_{n}|^2}}=0.\label{eqb}
\ee
Finally, we substract from Eq.(\ref{eqb}) its complex conjugate, obtaining
\be
{d\over{d t}}|C_{n}|^2=0
\ee
which implies $C_{n}(t)=C_{n}(0)=\delta_{n d}$. Thus, at $s\rightarrow 0$ there is complete trapping of the initial excitation, regardless of the impurity strength $\chi$. This is clearly seen in Fig.5. 
For finite $s$, the trapping curves show that for the surface case more impurity strength is needed to effect a similar trapped value as in the bulk case. Perhaps this is a manifestation of a sort of a position `uncertainty' effect: The presence of a surface confines the impurity much more than in the bulk case, thus increasing its kinetic energy; thus the need for a stronger `potential well' $\chi$ to
effect trapping.

{\bf Conclusions}.\\
We have examined analytically and numerically the physics of a single saturable impurity embedded in a 1D lattice, when the usual discrete Laplacian is replaced by a fractional one, characterized by a fractional exponent. We considered two types of impurity: a bulk one, located far away from the edges of the lattice, and a surface impurity located at one of the edges (`site zero'). By means of the formalism of lattice Green functions, we determined the existence of a single bound state. While for the bulk case there is a bound state for any amount of impurity strength, for the surface case a minimum amount of impurity strength is needed. These features are markedly different from the well-known case of a cubic impurity\cite{cubic,cubic2}. There, for the bulk case a bound state is only possible for impurity strengths larger than a critical value, while a surface impurity also requires a minimum impurity strength to generate a bound state, and up to two bound states are possible. Rather, our results resemble the ones for the linear impurity. The fractional exponent does not seem to play an important role in this respect.

The bound state energy curves are more or less similar for the bulk and surface cases, for given values of the fractional exponent. When this value is close to unity (standard case), the bound state spatial profiles resemble the ones found for the cubic impurity where the stagered-unstagered symmetry is obeyed. However as the exponent approaches zero, this symmetry is no longer obeyed.

The transmission of plane waves across the impurity decreases in an overall sense, since the energy interval of the passing waves shrinks with a decrease of the fractional exponent. In fact, in the limit of a vanishing exponent, the only wave that can be transmitted is the one with energy equal to $2$. 

The selftrapping of an initially-localized excitation is no dissimilar for the bulk and surface cases. In both cases it increases monotonically with an increase in impurity strength and, for a fixed impurity strength, it increases with a decreasing fractional exponent. The main difference between the two cases is that for the surface case, it takes more impurity strength to effect a degree of trapping. For both cases we observe the existence of `linear' trapping at small impurity strength, for all fractional exponents. In the limit of a vanishing exponent, the trapped fraction converges to unity, regardless of the impurity strength value. This can be traced back to the vanishing of the effective coupling $K^s(m)$  at small fractional exponents.

All in all, the main effect of fractionality in this saturable impurity was rather secondary, except at small values of the fractional impurity. This is the regime where the bandwidth is substantially shrunk, pushing all eigenvalues together and thus, inducing a tendency towards degeneration.

{\bf Acknowledgments}

This work was supported by Fondecyt Grant 1200120.

\end{document}